\documentclass[12pt]{article}
\usepackage{amsfonts,amssymb,amsmath,amsthm}
\usepackage{url}  

\newcommand{\nc}{\newcommand}
\newcommand{\comm}[1]{} 
\nc\drc{\DeclareRobustCommand}
\newcommand{\lb}[1]{\label{#1}}
\newcommand{\rf}[1]{(\ref{#1})}
\newcommand{\erf}[1]{Eq.\,\rf{#1}}
\newcommand{\be}{\begin{equation}}
\newcommand{\ee}{\end{equation}}
\newcommand{\el}[1]{\label{#1}\end{equation}}
\newcommand{\erl}[1]{\label{#1}\end{eqnarray}}
\def\Ga{\Gamma}

\def\spr#1{\langle #1 \rangle }
\newcommand{\for}{\quad\mbox{for}\quad}
\newcommand{\aand}{\quad \mbox{and}\quad}

\newcommand{\where}{\quad \mbox{where}\quad}

\nc\hence{\ifmmode{\curvearrowright}\else{\mbox{$\curvearrowright$\ }}\fi}%

\nc{\nin}{\noindent}
\def\edd{\end{document}}
\nc\bal{\begin{align}}

\newcommand{\Ra}{\ensuremath{\Rightarrow}} 
\nc{\tr}{\mbox{tr}\,}

\nc\bes{\begin{subequations}}
\nc\ees{\end{subequations}}
\def\N{\mathbb{N}} 
\def\Z{\mathbb{Z}} 
\def\x{\times}
\nc\cN{{\cal N}}\nc\cM{{\cal M}}
\def\cO{O}
\drc{\cA}{\ifmmode{\mathcal A}\else{\mbox{${\cal A}$\ }}\fi}%
\drc{\urb}[1]{{\noindent\cob{\url{#1}}}}
\nc{\bb}{{\bf b}}
\nc{\vb}{{\bf v}}

\newtheorem{theorem}{Theorem}  

\newtheorem{corollary}[theorem]{Corollary}
\newtheorem{lemma}[theorem]{Lemma}
\newtheorem*{example}{Example}   
\newtheorem{exercise}[theorem]{Exercise}

\newtheorem{proposition}[theorem]{Proposition}

\nc\bc{\begin{corollary}}
\nc\ec{\end{corollary}}
\nc\bex{\begin{exercise}}
\nc\eex{\end{exercise}}
\nc\bxml{\begin{example}}  \newcommand{\bexm}{\begin{example}}
\nc\exml{\end{example}}    \newcommand{\eexm}{\begin{example}}
\nc\blem{\begin{lemma}}
\nc\elem{\end{lemma}}
\nc\bpr{\begin{proposition}}
\nc\epr{\end{proposition}}
\newtheorem*{remark}{Remark}  
\nc\bre{\begin{remark}}    \nc\ere{\end{remark}}
\newtheorem*{definition}{Definition} 

\nc\bd{\begin{definition}}   
\nc\ed{\end{definition}} 
\nc\bp{\begin{proof}}
\nc\ep{\end{proof}}
\nc\bth{\begin{theorem}}
\nc\et{\end{theorem}}   

\begin{document}
\begin{titlepage}
\vskip 2cm
\begin{center}
\textbf{Generalized Orbits-Fixedpoints Relations}\\
\vskip 8pt
\textbf{Jamil Daboul}
\vskip 2pt
Physics Department\\
Ben Gurion University of the Negev\\
Beer Sheva, Israel \\
\vskip 2pt
{\tt E-mail:\ jdaboul@gmail.com}

\begin{abstract} 

I prove the following equality for $t$-transitive groups $G$
\[
\frac 1 {|G|} \sum_{g\in G} f_{\Z_N}(g)^k = \cN_{orbits} (G, \Z_N^k)
=  \sum_{j=1}^{\min (k,N)} d_j(G)\, S(k,j)\,, 
\]
where $S(k,j)$ are Stirling numbers of the second kind, and
$d_j(G) =1$ for $1\le j \le t$, and $d_j(G) \ge 2$ for $j > t$. 
The above equality extends further, using new proofs, two generalizations of an earlier  fixedpoints-orbits theorem for finite group actions $(S_N,\Z_N^k)$. 
An illustration using Mathieu group $M_{24}$ is discussed. Possible applications using tensor products of matrix permutation representations is indicated.

\end{abstract}
\end{center}
\emph{Keywords:} \texttt{group action, orbits, fixed points, Burnside lemma, multiply transitive finite groups, Bell numbers, Stirling numbers of the second kind, symmetry group $S_N$, Mathieu groups, permutation representations}. \\

MSC-class: 20B20 (primary); 20Cxx, 20C35 (secondary)
\end{titlepage}

\section{Introduction}

In 1975 Goldman \cite{goldman} proved the following equality, by using interesting statistical (!) arguments:

\begin{theorem} \label{Goldman} The following equality holds for the 
symmetric group $S_N$: 
\be
\spr{f_{\Z_N}^k} :=\frac 1 {|S_N|} \sum_{g\in S_N} f_{\Z_N}(g)^k= \sum_{j=1}^{\min (N,k)} S(k,j)\,, \for k, N\ge 1 \,,
\el{sn}
where $S(k,j)$ are the Stirling numbers of the second type.
\end{theorem}

\bd
The  \textbf{Stirling numbers of the second kind $S(k,j)$ }
count the number of ways to partition the set $\Z_k=\{1,2,\ldots,k\}$ into $j$ nonempty subsets. For example, the three integers in $\Z_3=\{1,2,3\}$ can be 
separated into $j=2$ subsets in $3=S(3,2)$ different ways: 
$$
\{\{1\},\{23\}\},~\{\{2\},\{13\}\},~\{\{3\},\{12\}\}.
$$
The $S(k,j)$ can be calculated by using their \textbf{generating function}  \cite{stirling}:
\be
  x^k=  \sum_{j=0}^k S(k,j) (x)_j 
  = S(k,0) + \sum_{j=1}^k S(k,j)\, x(x-1)\cdots(x-j+1)\, . 
\el{gf}
where $k\in \N$ and $(x)_j$ is the \textbf{falling factorial}, with $(x)_0=1$. 
It follows from \rf{gf} that $S(k,0)= \delta_{k,0}$. Hence, by substituting $x=N\in\N$ in \rf{gf} we obtain for $k\ge 1$:  
\be
   N^k = \sum_{j=1}^k S(k,j) (N)_j \,, ~~~\forall~~ N, k\ge 1.  
\el{gN}
where 
\be
(N)_j := N(N-1)\ldots (N+1-j)= \frac{N!}{(N-j)!}
, \for 1\le j\le N\,.
\el{nj}
Note that $(N)_j:=0$ for $j > N$, and $(N)_{(N-1)}=(N)_N=N!$.
\ed

For $k \le N$ \erf{sn} reduces to \cite{huppert,van Lint}:
\be
\spr{f_{\Z_N}^k} :=\frac 1 {|S_N|} \sum_{g\in S_N} f_{\Z_N}(g)^k= B_k\,, \for N\ge k\ge 1 \,,
\el{kn}
where $B_k$ denotes the Bell numbers which are related to $S(k,j)$ by \cite{stirling} 
\be
B_k :=  \sum_{j=1}^{k} S(k,j)\,, \for k \ge 1 \,.
\el{bk}
A second generalization of \erf{kn} was obtained by extending the validity of \rf{kn} to general $t$-transitive groups $G$ instead of just the symmetric group $S_N$:

\bth \emph{\cite{merris,mt}}\lb{merris} 
G is t-transitive on X, if and only if
\be
\frac 1 {|G|} \sum_{g\in G} f_{\Z_N}(g)^k = B_k\, ,  \for k\le t\le N:=|X|\,.
\el{mon}
\et 
The equality \rf{mon} was first proved by Merris and Pierce (1971) \cite{merris} by induction on $k$. 
A second proof was given by Monro and Taylor (1978) \cite{mt},  by mapping subsets of $\Z_N^k$ onto partitions of $\Z_k$.

\vspace{0.15in}
In present paper I give in theorem \ref{myth} a new proof of \rf{sn}, based on Burnside lemma \cite{bw}. My proof also shows that the r.h.s. of \rf{sn} is equal to the number of orbits of the action $(S_N, \Z_N^k)$; this interpretation cannot be deduced from the proof of Goldman \cite{goldman}. This interpretation is important, since it enables me to give a simple proof of \erf{dab} below, which is valid for any finite group $G$ and also for $k>N$. 

\section{New proofs and results}
\bd
When a group $G$ acts on a G-set Y, it decompose it into disjoint orbits.
In particular, when $S_N$ acts on $Y=\Z_N^k$, it produces \textbf{$S_N$-orbits} of different types, as follows:
\be 
\cO_{j,k,N}:= S_N\cdot \bb_{jk} \in \Z_N^k, ~~j \le \min(k,N)\,,
\el{okjn}
where $\bb_{jk}$ denotes \textbf{basis ordered}  $k$-tupels which depend on $j$ distinct integers from $\Z_j=\{1,2,\ldots,j\}$ (not from $\Z_N$) 
\[
\bb_{jk}:= (x_1, \ldots, x_k) \in \Z_j^k, \where x_i\in \Z_j\,.
\]
\ed

\bxml
To illustrate the above notation, consider an $S_N$-orbit: 
\bal 
\cO_{3,4,N}&:= S_N\cdot \bb_{34}=S_N\cdot (1,2,3,2)
= \{(g(1),g(2),g(3),g(2))| g\in S_N\}\cr
&= \{(i,j,k,j)| i\ne j\ne k\ne i \in \Z_N\}\,. \nonumber
\end{align}
\exml

\bth \label{cor} The number of orbits created by the group action $(S_N,\Z_N^k)$ is given  by
\be
\cN_{orbits} (S_N,\Z_N^k) = \sum_{j=1}^{\min (k,N)} S(k,j)\,, \for k, N\ge 1 \,, \el{osn}
\comm{
\bes \lb{osn}
\bal
\cN_{orbits} (S_N,\Z_N^k) &=\frac 1 {|S_N|} \sum_{g\in S_N} f_{\Z_N}(g)^k \lb{osn1}\\
&= \sum_{j=1}^{\min (N,k)} S(k,j)\,, \for k, N\ge 1 \,, \lb{osn2}
\end{align}
\ees \endcomm}
\et
\bp 
\comm{The first equality in \erf{osn1} follows from \rf{fx}, which is valid for any 
group $G$. 
The second equality in \erf{osn2} follows directly from \erf{sn}.
But I shall now give a new proof of \erf{sn}, as follows:\endcomm}
The length of an orbit $\cO_{j,k,N}$ is independent on $k$:
\be 
|\cO_{j,k,N}|=|\cO_{j,j,N}|= (N)_j:=N(N-1)\cdots(N+1-j), ~~j \le N\,,
\el{}
Let $n(k,j)$ denotes the number of orbits of type $\cO_{j,k,N}$. It follows that
\be 
N^k = \sum_{j=1}^{\min(k,N)} n(k,j) |\cO_{j,k,N}|
    = \sum_{j=1}^{\min(k,N)} n(k,j) (N)_j\,.
\el{nkn}
By comparing \erf{nkn} with the equality \erf{gN}, starting from $N=1$ and successively $N=2,3,\ldots$, we obtain 
\be
n(k,j)=S(k,j), \quad \forall~~ k\ge j\ge 1\,. 
\el{nes}
\ep

\blem \label{lem} Let $f_{X}(g)$ and $f_{X^k}(g)$ denote the number of fixed points of the actions $(G,X)$ and  $(G,X^k)$, respectively. Then
\be   
\spr{f_X^k} = \spr{f_{X^k}} = \cN_{orbits} (G,X^k)\,.
\el{fx}
\elem

\bp
We recall that $f_{X^k}(g)$ is equal to the number of ordered k-tupels $(x_1,\ldots,x_k) \in X^k$ which are fixed by the action of $g\in G$. Hence, the first equality in \rf{fx} follows from 
\begin{align}
f_{X^k}(g) &= \sum_{(x_1,\ldots, x_k) \in X^k} \delta_{(x_1,\ldots,x_k)\,,~ g\cdot\,(x_1,\ldots, x_k)} \cr 
&= \prod_{j=1}^k (\sum_{x_j \in X} \delta_{x_j\,,~ g\cdot x_j})
= f_X(g)^k\,, \lb{xkk}
\end{align}
since the summations over $x_j$ can be carried out independently. 
The second equality follows immediately 
from Burnside's lemma \cite{bw} for $(G,X^k)$. 
\ep

Below I generalize the two equalities \erf{sn} and \erf{mon}, as follows:

\bth \label{myth} The group action $(G,\Z_N)$ is t-transitive, if and only if the following equality holds:
\bes \lb{dab}
\bal
\frac 1 {|G|} \sum_{g\in G} f_{\Z_N}(g)^k 
&= \cN_{orbits} (G, \Z_N^k)  \lb{dab1} \\
&=  \sum_{j=1}^{\min (k,N)} d_j(G)\, S(k,j)\,,  \lb{dab2} \\[2.5mm]
&\Ra \left\{\begin{array}{ll}
 B_k\,, &for~ k\le t\le N\,, \nonumber\\[2.5mm]
 B_t + \sum_{j=t+1}^{\min (k,N)} d_j(G) S(k,j)&for~ t< k\,, 
 \end{array} \right .
\end{align}
\ees
where the divisions $d_j(G)$ depend on the group $G$ and on $j$, but not on $k$:
\be
\begin{array}{lll}
d_j(G) &=1      &for\quad   1\le j \le t, \aand \\[2mm]
d_j(G) &\ge 2   &for\quad  j > t\,. 
\end{array} 
\el{dj}
\et

\bp
\erf{dab1} follows from \erf{fx}.\\

\erf{dab2} follows from \erf{osn} after taking into account: 
\begin{itemize}
	\item Every $G$ which acts on $\Z_N$ must be a subgroup of $S_N$.
\item If $G$ is a genuine subgroup of $S_N$, it would have less group elements. Therefore, we expect 
\be
|G\cdot \bb_{jk}| \le |S_N \cdot \bb_{jk}|\,. 
\el{length}

\item The equality sign in \rf{length} ($d_j(G)=1$) holds
for $j\le t$, since  a t-transitive group produces exactly the same orbits $\cO_{j,k,N}$, as $S_N$, for $j\le t$. 

\item For $j>t$ then $|G\cdot \bb_{jk}| < |S_N \cdot \bb_{jk}|$, which means that 
the corresponding (maximal) orbit $\cO_{j,k,N}$ of $S_N$ must split into $d_j(G)\ge 2$ orbits of $G$.
\end{itemize}
\ep

\bxml
The Mathieu group $M_{24}$ is 5-transitive. Its divisions $d_j$ can be read from the following formula. \emph{(Note that since $S(k,j)=0$ for $k < j$, the formula \rf{djm24} is valid for all $k\ge 1$.)}
\begin{align} 
\spr{f^k_{\Z_{24}}} &= \sum_{j=1}^{5} S(k, j) + 2\, S(k, 6) + 9\, S(k, 7) 
+ 123\, S(k, 8)\cr 
     & \qquad + 1938 \sum_{j=9}^{\min(k,24)} \frac{15!}{(24 - j)!} S(k, j)\,. \lb{djm24}
   \end{align}
We can easily understand why the maximal $S_{24}$-orbit $\cO_{6,6,24}$ has to split:
$|\cO_{6,6,24}|= (24)_6= (24)_5\cdot 19$ is not a divisor of $|M_{24}|= (24)_5\cdot 48$. Hence, I conclude that $\cO_{6,6,24}$ must split into two sub-orbits, with lengths $(24)_5\cdot 16$ and  $(24)_5\cdot 3$, which are divisors of $|M_{24}|$. 
Note that $d_7=9$ and $d_8=123$ yield more predictions.
\exml

\section{Summary} 
 
I gave a new proof of \erf{sn} by using the generating function of $S(k,j)$
and Burnside lemma. Unlike the statistical proof of Goldman \cite{goldman}, my proof
led to the equality \rf{dab1}, between the r.h.s. of \rf{sn} to the number of orbits of $(S_N,\Z_N)$. This made it possible to derive \erf{dab}, which is a generalization
of \rf{mon} to $k>N$, by using a simple argument which led to the inequality in \erf{length}.

Note that a finite group action $(G,\Z_N)$ yields an $N\x N$-matrix representation of $G$, which is called \textbf{permutation representation} $\Ga^P$, so that the action $(G,\Z_N^k)$ yields a k-fold tensor product of $\Ga^P$.  

A detailed version of the present paper is available as a preprint, which includes applications and a review of basic concepts of group action.
I will gladly send it by email upon request.

\edd